\documentclass[pra,twocolumn,amsmath,amssymb,showpacs,floatfix]{revtex4}

\usepackage{graphicx}
\usepackage{hyperref}

\let\origcite\cite
\def\cite#1{\unskip~\origcite{#1}}
\newcommand{\refe}[1]{\unskip~(\ref{#1})}

\newcommand{\etal}{{\em et al.}}

\renewcommand{\vec}[1]{\mathbf{#1}}
\newcommand{\hv}[1]{\hat{\mathbf{#1}}}

\newcommand{\sbsqrt}[1]{\sqrt{\smash[b]{#1}}} 

\usepackage{multirow}
\newcommand{\MeijerG}[5]{%
G^{#1}_{#2}
\def\arraystretch{0.8}
\biggl(
\begin{array}{l|l}
\multirow{2}{*}{\!#3}  
& #4 \\[2pt]
& #5
\end{array}\!
\biggr)
}

\begin{document}

\title{On cold gases with anisotropic interactions}
\author{Alexander Pikovski}
\affiliation{
Zentrum f\"ur Optische Quantentechnologien, Universit\"at Hamburg, 22761 Hamburg, Germany
}
\altaffiliation{Present address: JILA, NIST and Department of Physics, University of Colorado, Boulder, CO 80309, USA.}
\date{\today}
\begin{abstract}
A cold gas of particles with anisotropic interactions of general form, due to a polarizing field, is studied.
Special cases are atoms or molecules with dipole--dipole or quadrupole--quadrupole interactions.
It is shown that the angular dependence of an observable on the direction of the polarizing field is largely determined
by symmetry. For a gas in a confined quasi two-dimensional geometry, the effective interaction is calculated in general
form. Some examples of dipole and quadrupole gases are considered. 
It is concluded that when anisotropic forces are studied in a general manner, one can obtain
simpler results and better understanding for some problems.
\end{abstract}

\pacs{%
67.85.-d, 
33.15.Kr
}

\maketitle

\section{Introduction}

The creation of cold quantum gases with dipole moments 
has stimulated studies of systems with anisotropic interparticle
interactions. Atomic gases with magnetic dipole moments
\cite{Lahaye-2007,Lu-2011,Aikawa-2012,Beaufils-2008}
and diatomic molecules with electric dipole moments
\cite{Ni-2008,Deiglmayer-2008}
are being studied experimentally.
Theoretical investigations have been concerned with many aspects of the physics of dipolar
gases, see \cite{Lahaye-2009,Baranov-2012} for an overview. It has also been proposed to produce
cold gases with electric quadrupole moments \cite{Derevianko-2001}.

The forces between charge dipoles or magnetic dipoles have anisotropic form, 
that is, the interaction between two particles depends not only on the distance but also on the
angles between the particles. 
Such anisotropic interactions have attracted great interest since they may lead to novel many-body phenomena 
(see overview in \cite{Baranov-2012}).
The main feature of these proposals is the angular dependence of the interaction.
Also, schemes have been proposed to modify the angular dependence of the dipole interactions,
by using additional fields \cite{Baranov-2012,Gorshkov-2013,Li-2013b} or optical lattices \cite{Wall-2013}.
Quadrupole interactions in cold gases have received some attention recently 
\cite{Bhongale-2013,Olmos-2013,Li-2013a}.

In this paper, we consider a cold gas of particles which are polarized by an external field 
and interact by anisotropic forces of quite general form (Sec.~\ref{sec-aniso}),
dipole--dipole and quadrupole--quadrupole interactions being special cases.
It is shown how the dependence on the direction of the polarizing field can be
found on the basis of symmetry (Sec.~\ref{sec-geometry}). 
For a quasi two-dimensional geometry,
the effective interaction and its Fourier transform are calculated in general 
form (Sec.~\ref{sec-q2d}). 
As an example, the ground-state energy of a gas with anisotropic interactions 
is calculated for a layer geometry and for a lattice (Sec. \ref{sec-examples}).
It is concluded that some effects due to anisotropic forces can be conveniently studied in a general way, 
without specifying the exact form of the interactions.

\section{Anisotropic interactions}
\label{sec-aniso}

\begin{table*}[th]
\vspace{-6pt}
\begin{ruledtabular}
\begin{tabular}{lll}
Symmetry & Point group & Invariant spherical harmonics 
\vspace*{3pt}
\\\hline
\noalign{\vskip 3pt}
ellipsoid with axes $a \ne b \ne c$;
rectangular box & $D_{2h}$ & $Y^C_{\ell,m}$ with $l$ even, $m$ even\\
square prism & $D_{4h}$ & $Y^C_{\ell,m}$ with $\ell$ even, $m$ multiple of $4$\\
full cubic symmetry & $O_h$ & Cubic harmonics, formed from \\
& & $Y^C_{\ell m}$ with $m$ multiple of $4$, for $\ell=0,4,6, \ldots$\\
rotational symmetry (in $xy$ plane) & $SO(2)$ & $Y_{\ell m}$ with  $m=0$
\end{tabular}
\end{ruledtabular}
\vspace{-6pt}
\caption{Common symmetry types and spherical harmonics which form an invariant basis.
For cubic harmonics, see e.g.\ \cite{Mueller-1966}.
}\label{table-1}
\end{table*}

The interaction between two atoms or molecules depends on the states of the particles and can be
exceedingly complicated. Here we consider the simple case where the interaction between two particles
at positions $\vec{r}_1$ and $\vec{r}_2$
can be represented by a scalar
potential function $V(\vec{r})$, which however may be anisotropic, i.e. it depends
on the direction of the distance vector $\vec{r}=\vec{r}_1-\vec{r}_2$.
The interaction potential of atoms or molecules, which are polarized by an external field 
${\vec{F}}$, can be often written as
\begin{equation}\label{Vgen}
V(\vec{r}) = \sum_{n} A_n V_n(|\vec{r}|) P_n(\hat{\vec{r}}\cdot \hat{\vec{F}})  
\end{equation}
where hats indicate unit vectors and $P_n(x)$ is the Legendre polynomial. 
Examples of interactions of such form will be discussed in the following, see also \cite{OMalley1964}.
If parity is present, i.e.\ $V(\vec{r})=V(-\vec{r})$, then
the sum in Eq. \refe{Vgen} runs only over even $n$.

The interaction of two atoms or molecules has often the form of the interaction of
classical charge distributions. 
A quantum-mechanical calculation of first-order interactions gives
the classical interaction of electronic charge densities \cite{Hirschfelder-Meath};
the same form, only with a modified prefactor, is also obtained
in other cases such as resonant interaction,
see e.g.\ \cite{Hirschfelder-Meath,Byrd,Pikovski-2011,Gorshkov-2011}.
The interaction energy of two (non-overlapping) classical charge distributions can be written in the form of a multipole
expansion $V=\sum_n C_n/r^n$. Each $C_n$ contains contributions of the different
electric multipoles of the charge distributions (the interaction between a $2^n$-pole and a $2^m$-pole has
a $1/r^{n+m+1}$ dependence), with certain angular dependencies. 
Assuming a rotational symmetry with respect to the axis of $\hv{F}$, however, only interactions of the form \refe{Vgen}
are possible.
Expressions such as Eq. \refe{Vgen} are only valid at distances much larger than the atom size,
and for small $r$ some care is required, e.g.\ for dipoles see \cite{Andreev-2013,Eberlein-2005}.

In the following, we will be interested in how some observables depend on the direction of the applied field.
The angular dependence associated with the direction of $\vec{F}$ can be disentangled in Eq. \refe{Vgen}
with the help of the addition theorem of spherical harmonics:
\begin{equation}\label{add-thm}
\begin{split}
& \hspace{-1em} P_\ell(\hv{r}\cdot\hv{F}) = \frac{4\pi}{2\ell+1} \!\sum_{m=-\ell}^\ell \!
Y_{\ell m}^\ast (\hv{r}) Y_{\ell m} (\hv{F}) \\[3pt]
= & P_\ell(\hv{r}\cdot\hv{z}) P_\ell(\hv{F}\cdot\hv{z}) \\
& + \frac{8\pi}{2\ell+1} {\displaystyle \sum_{m=1}^\ell  }
\Bigl( Y_{\ell m}^C (\hv{r}) Y_{\ell m}^C (\hv{F}) +  Y_{\ell m}^S (\hv{r}) Y_{\ell m}^S (\hv{F}) \Bigr) 
\end{split}
\end{equation} 
with the notation 
$Y_{lm}(\hv{x})=Y_{lm}(\theta,\phi)$ where $(\theta, \phi)$ are the angles of $\hv{x}$
in spherical coordinates. Here $Y_{lm}^C=\text{Re}\, Y_{lm}$ and $Y_{lm}^S=\text{Im}\, Y_{lm}$ are the
real spherical harmonics with polar angular dependence $\cos(m\phi)$ and $\sin(m\phi)$ respectively.

It is often required in calculations to obtain the Fourier transform of an expression of the type as in Eq. \refe{Vgen}.
It can be found using the identity \cite{Bochner}:
\begin{equation}\label{Bochner3D}
 \int \!\! f(|\vec{r}|) Y_{\ell m}(\hv{r}) e^{i \vec{k} \cdot \vec{r}} d^3r
= 4\pi i^\ell \, Y_{\ell m}(\hv{k}) \! \int_0^\infty \!\! t^{2} f(t) j_{\ell}(t |\vec{k}|) dt,
\end{equation}
where $j_\ell(x)$ is the spherical Bessel function.
For power-law potentials, one has $f(r)=1/r^p$, and we note that 
$
 \int_0^\infty \!\! t^{2-p} j_{\ell}(kt) dt = k^{p-3} \frac{\sqrt{\pi}}{2^{p-1}}  \frac{\Gamma((3+\ell-p)/2)}{\Gamma((\ell+p)/2)},
$
for $p>1$ and $p<3+ \ell$. 

\section{Influence of geometry}
\label{sec-geometry}

Cold gases in optical traps can be confined to different geometries by modifying the trap shape and 
applying optical lattices. 
An observable, which is affected by anisotropic interparticle interaction, depends on the direction
of the external field $\hv{F}$ not arbitrarily, but in a way restricted by the symmetry of the sample. 

To be concrete, consider the interaction energy of a many-body system. 
It can be expressed in the form
\begin{equation}\label{U}
 U = \frac{1}{2} \int V(\vec{r}_1-\vec{r}_2) g(\vec{r}_1,\vec{r}_2) \,d^3r_1 d^3r_2 .
\end{equation}
The distribution function $g(\vec{r}_1,\vec{r}_2)$
is the probability to find one particle at $\vec{r}_1$ and another one at $\vec{r}_2$,
it is given in second quantization by
$g(\vec{r}_1,\vec{r}_2) = \langle \psi^\dag(\vec{r}_1) \psi^\dag(\vec{r}_2) \psi(\vec{r}_2) 
\psi(\vec{r}_1) \rangle$.
In a cold dilute gas 
it is often sufficient to treat the effects of anisotropy to first order.  
(Effects such as Fermi-surface deformation, reviewed in \cite{Baranov-2012}, are of second order.)
This means that $g(r_1,r_2)$ can be taken as that of the gas with isotropic interactions only.
One can then make use of symmetries of the system to find the possible dependence of $U$ on the angle of $\hv{F}$.
To simplify the discussion even further, let us consider a particle at the origin of 
coordinates. Its energy due to the anisotropic interaction $V=P_\ell(\hv{r}\cdot\hv{F})$ is then proportional to
(using Eq. \refe{add-thm}):
\begin{equation}\label{Uexp}
\! \int \! V(\vec{r}) g(0,\vec{r}) d^3r \sim \!\!\sum_{m=-\ell}^\ell \! Y_{\ell m}^\ast(\hv{F}) 
\!\! \int \!\! Y_{\ell m}(\hv{r}) V_\ell(|\vec{r}|) g(0,\vec{r})  d^3r .
\end{equation}
If the system is invariant under some symmetry transformations,
the integral on the right-hand side of Eq. \refe{Uexp} must vanish identically for some $Y_{\ell m}$. 
To be more precise, consider a system which is invariant with respect to a group of transformations 
which leave the origin fixed, i.e.\ a point group.
One can form linear combinations of spherical harmonics for fixed $\ell$ which form
an invariant basis for this point group and compare with the right-hand side of Eq. \refe{Uexp} 
[or Eq. \refe{add-thm}] to see which harmonics must vanish. 
Table \ref{table-1} lists some common symmetries and the corresponding invariant bases composed of $Y_{\ell m}$
or $Y^C_{\ell m}$. Examples will be given in Sec.\ \ref{sec-examples}.


\section{Quasi two-dimensional geometry}
\label{sec-q2d}

Quasi two-dimensional geometry, meaning that the gas is confined to a thin layer, 
is a particularly interesting arrangement, since it allows the study of two-dimensional physics.
Let us consider the gas to be confined in the $z$ direction to a thin layer in the $xy$ plane. 
The many-body Hamiltonian will be of the form $H=K+U+V_c$, where $K$ is the kinetic energy, $U$ the internal 
energy of Eq. \refe{U} and $V_c$ the confinement potential.
If all wave functions can be written as
\begin{equation}
 \psi(x,y,z) = \psi(x,y) \chi(z),
\end{equation}
then $g(\vec{r}_1,\vec{r}_2)=g_{2d}(x_1y_1,x_2y_2)\chi^2(z_1)\chi^2(z_2)$.
Changing to relative and center-of-mass coordinates, we see
that the internal energy [Eq. \refe{U}] can be presented in the form of a two-dimensional integral
\begin{equation}\label{U2d}
 U = \frac{1}{2}\int V_{q2d} (\vec{s}_1-\vec{s}_2) g_{2d}(\vec{s}_1,\vec{s}_2) \, d^2s_1 d^2s_2
\end{equation}
where
\begin{equation}\label{Vq2d-gen}
 V_{q2d}(x,y) \! = \!\! \iint_{-\infty}^\infty \!\!\!\! V(x,y,z) \chi^2\bigl(Z\!+\!\frac{z}{2}\bigr) \chi^2\bigl(Z\!-\!\frac{z}{2}\bigr) \, dz dZ .
\end{equation} 
For sufficiently deep confinement potential, one can consider the confinement potential to be harmonic.
For a particle in the ground state of the harmonic potential,
the wave function in the $z$ direction is
\vspace{-6pt}
\begin{equation}
\qquad 
\chi(z) = \frac{1}{(\sqrt{\pi} w)^{1/2}} \exp\Bigl(- \frac{ z^2 }{ 2 w^2} \Bigr) ,
\vspace{-3pt}
\end{equation} 
where $w$ is the oscillator length of the harmonic oscillator,
and Eq. \refe{Vq2d-gen} becomes
\vspace{-6pt}
\begin{equation}\label{Vq2d}
 V_{q2d}(x,y) =  \frac{1}{\sqrt{2\pi} w} \int_{-\infty}^\infty \! V(x,y,z) \exp\Bigl(-\frac{z^2}{2w^2}\Bigr) \, dz .
\vspace{-3pt}
\end{equation}
The potential $V_{q2d}$ can be viewed as an effective two-dimensional potential.
The many-body hamiltonian acquires a strictly two-dimensional form.
This two-dimensional description applies only as long as the physics at length scales much smaller that $w$ is
not probed, however.

\begin{figure}[t]
\vspace{-0.6cm}
\centering
\includegraphics[width=0.9\columnwidth]{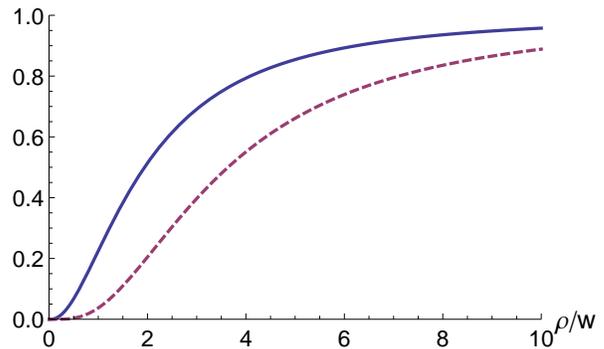}
\vspace{-6pt}
\caption{Plot of the function $F(\rho/w)$ for $n=2$ (solid) and $n=4$ (dashed).}\label{fig}
\vspace{-0.5cm}
\end{figure}

Let us discuss the effective two-dimensional potential for an interaction of the form
\vspace{-6pt}
\begin{equation}
V_n(\vec{r}) = \frac{1}{r^{n+1}} \frac{P_n(\hv{r}\cdot\hv{F})}{P_n(0)}  .    
\vspace{-3pt}
\end{equation}
This potential appears e.g.\ as the 
dipole--dipole ($n=2$) or quadrupole--quadrupole ($n=4$) interaction, see Sec.~\ref{sec-examples}.
The resulting two-dimensional effective potential, Eq. \refe{Vq2d}, can be calculated in closed form. 
The integral one needs to calculate contains spherical harmonics of
different order $m$, cf.\ Eq. \refe{add-thm}. To illustrate the main features, we will discuss the $m=0$ in detail, 
calculations for other orders can be done similarly. 
The effective two-dimensional potential for $m=0$ is
\begin{equation}
 V_{q2d}^{(n)}(\rho) =  \frac{1}{P_n(0)\sqrt{2\pi} w} \int_{-\infty}^\infty \! 
\frac{ 
P_n\Bigl(\!\frac{z}{ \sqrt{\rho^2+z^2} }\!\Bigr) 
e^{-\frac{z^2}{2w^2}}
}{(\rho^2+z^2)^{(n+1)/2}} 
\, dz
\end{equation}
where we have used $z=r\cos\theta$ and $\rho=\sqrt{x^2+y^2}$. 
Writing $t=z/\rho$ and $u=\rho/w$, this is $I_n^0 / (P_n(0)\rho^{n+1}) $, where $I_n^0$ is the integral calculated in the Appendix.
Therefore we have, for even $n$,
\vspace{-6pt}
\begin{equation}\label{Vq2dn}
 V_{q2d}^{(n)} (\rho) = \frac{1}{\rho^{n+1}} F(\rho/w)
\vspace{-3pt}
\end{equation}
with 
\begin{equation}\label{F}
 F(x) = \Bigl(\frac{x^2}{2}\Bigr)^{(n+1)/2} \Psi\left(\frac{n+1}{2},1,\frac{x^2}{2}\right)
\end{equation}
where $\Psi$ is the Tricomi confluent hypergeometric function.
The function $F(x)$ behaves as $F(x) \sim x^{n+1}$ for $x\to 0$ and approaches $F(x)=1$ for $x\to \infty$,
a plot of $F(x)$ is shown in Fig. \ref{fig}. 
Thus for distances much larger than $w$ the potential in Eq. \refe{Vq2dn} approaches
a strictly two-dimensional form $V(\rho)=1/\rho^{n+1}$, as expected.

The two-dimensional Fourier transform of the potential in Eq. \refe{Vq2dn} is 
$V_{q2d}^{(n)}(k) = 2\pi \int_0^\infty \!\! J_0(k \rho) V_{q2d}^{(n)}(\rho) \rho d\rho$.
It can be calculated by a method similar to
the one used in the Appendix, the result is:
\vspace{-6pt}
\begin{equation}\label{Vq2dk}
 \! V_{q2d}^{(n)}(k) = \frac{\pi}{\Gamma\bigl(\frac{1+n}{2}\bigr)} \Bigl(\frac{k}{2}\Bigr)^{n-1} 
\!\!\Psi\Bigl( \frac{1+n}{2},\frac{1+n}{2}, \frac{k^2w^2}{2} \Bigr)
\vspace{-3pt}
\end{equation}
It is helpful to note that
$V_{q2d}^{(n)}(k=0)=
a_n/w^{n-1}$, where 
$a_n=
2^{(3-n)/2} \pi [(n-1)\Gamma(\frac{n+1}{2})]^{-1}$.

The results \refe{Vq2dn} and \refe{Vq2dk} for $n=2$ (dipole--dipole interaction) are well-known
\cite{Fischer-2006,Babadi-2011,Baranov-2012}.
We note that the function in Eq. \refe{F} can be represented in terms of modified Bessel functions
and the function in Eq. \refe{Vq2dk} in terms of error functions.

\vspace{-6pt}
\section{Examples}
\vspace{-6pt}
\label{sec-examples}

As an example of the preceding discussion, we consider a gas of ultra-cold particles interacting by
dipole--dipole or quadrupole--quadrupole forces. 

The dipole--dipole forces have been reviewed in \cite{Lahaye-2009,Baranov-2012}. 
The interaction between two electric dipoles, both dipole moment vectors pointing along $\hv{F}$
and with magnitudes $d_1$ and $d_2$, is given by 
\vspace{-6pt}
\begin{equation}
V_D = \frac{1}{4\pi\epsilon_0} \frac{d_1 d_2}{|\vec{r}|^3} \frac{P_2(\hv{r}\cdot\hv{F})}{P_2(0)}
\vspace{-3pt}
\end{equation}
and $P_2(0)=-1/2$.

The interaction between two electric quadrupoles has been treated in the literature,
e.g.\ \cite{OMalley1964, Hirschfelder-Meath} and references therein.
For an axially symmetrical charge distribution there is only one number which characterizes the quadrupole moment.
Then, if the principal $z$ axis of quadrupole tensor is taken to point along $\hv{F}$, one can speak of the
$zz$ component of this traceless tensor as the quadrupole moment $Q$.
In this setting the interaction between two particles located at $\vec{r}_1$ and $\vec{r}_2$ 
with quadrupole moments $Q^{(1)}$ and $Q^{(2)}$ takes the form
\vspace{-6pt}
\begin{equation}\label{VQ}
V_Q = \frac{1}{4\pi\epsilon_0}\frac{3}{2} Q^{(1)} Q^{(2)} \frac{P_4(\hv{r}\cdot\hv{F})}{|\vec{r}|^5} 
\vspace{-3pt}
\end{equation}
where $\vec{r}=\vec{r}_1-\vec{r}_2$ and $P_4(0)=3/8$. 
The direct, first-order interaction between atoms has the
same form as the classical interaction between two quadrupoles, Eq. \refe{VQ},
only that $Q$ is now the expectation value of the operator 
$3z^2-r^2
= 2 r^2\sqrt{\frac{4\pi}{5}}Y_{2,0}(\theta)$ multiplied by the charge, where $z=r\cos\theta$.
For an atom or molecule in a state with 
total angular momentum quantum numbers $(J,M)$ with respect to an axis along $\hv{F}$,
this expectation value depends on the state.
It is conventional  to list the quadrupole moment for the state $(J,M \!=\!J)$.
The moment in a state with a different value of $M$ is \cite{OMalley1964}
$Q_{J,M}=\frac{3M^2-J(J+1)}{J(2J-1)} Q_{J,J}$, for $J \ge 1$
(Ref. \cite{Bhongale-2013} lists a different, possibly incorrect, expression for the quadrupole--quadrupole interaction).


\vspace{-8pt}
\subsection{Infinite homogeneous layer}
\vspace{-6pt}

Consider a gas of $N$ particles in a quasi two-dimensional geometry with wave functions being Gaussians in the $z$ direction, 
as discussed in Sec. \ref{sec-q2d}.
Let the interaction between particles be of the form
$V=\frac{C_{n+1}}{r^{n+1}} \frac{P_n(\hv{r}\cdot\hv{F})}{P_n(0)}$, for even $n$.
The ground-state energy shift per particle is $E=U/N$, which in view of the discussion of Sec. \ref{sec-q2d}
and Eq. \refe{U2d} is
\vspace{-6pt}
\begin{equation}\label{E2d}
 E = C_{n+1} P_n(\hv{F}\cdot\hv{z}) \cdot \frac{n_{2d}}{2} \int V_{q2d}^{(n)}(\rho) g(\rho) \,d^2r  ,
\vspace{-3pt}
\end{equation} 
where $g(\rho_1-\rho_2)=g_{2d}(\rho_1-\rho_2)/n^2_{2d}$ is the pair distribution function and $n_{2d}$ is the two-dimensional number density of particles. 
In Eq. \refe{E2d}, only the $m=0$ term remained in the 
angular expansion of Eq. \refe{add-thm},
since $g(r)$ for the noninteracting layer is isotropic (cf.\ Table \ref{table-1} with symmetry $SO(2)$).
For a two-dimensional Fermi gas which is spin-polarized, i.e.\ there is only one spin component, the pair distribution function at zero temperature is given by
\vspace{-6pt}
\begin{equation}\label{E-2d-g}
 g(\rho) = 1 - \frac{4}{k_F^2} \frac{J_1^2(k_F \rho)}{\rho^2} ,
\vspace{-3pt}
\end{equation} 
using $n_{2d} = k_F^2/(4\pi)$. 
This expression, when inserted in Eq. \refe{E2d}, contains the direct and exchange terms in the Hartree-Fock approximation.
Note that no short-distance cutoff is needed in the integral of Eq. \refe{E2d} for the dipole and quadrupole interactions, 
since $V_{q2d}^{(n)}(\rho)$ vanishes at the origin. 
(In a discussion of the dipole potential in Refs.\ \cite{Chan-2010,Kestner-2010} a purely two-dimensional potential $1/\rho^3$ was used, 
and a short-distance cutoff, of the order of the layer thickness $w$, was required. 
The function $F(\rho/w)$ provides automatically a cutoff here, in the same way as in Ref.\ \cite{Babadi-2011}.)
The energy shift $E$ is numerically small, because on the one hand $V_{q2d}$ vanishes for small distances and on the 
other hand the pair distribution function $g(\rho)$ for like fermions is zero for small distances.

For higher temperatures above quantum degeneracy, we may neglect the exchange term in $g(\rho)$
and put simply $g=1$. 
Then, using the result for $V_{q2d}^{(n)}(k=0)$ in Sec. \ref{sec-q2d}, we get
 $E= C_{n+1}\frac{n_{2d}}{2 w^{n-1}} a_n P_n(\hv{z}\cdot\hv{F}) $ where $a_n$
is given at the end of Sec. \ref{sec-q2d}.

\vspace{-8pt}
\subsection{Particles in a lattice}
\vspace{-3pt}

For a second example, let us consider atoms with dipole or quadrupole interactions which form a lattice.
We take them to be sufficiently well-localized at the lattice sites, one atom per site for simplicity, 
such that the interaction energy per particle $E$ of the particle at the origin is the sum
\vspace{-6pt}
\begin{equation}
 E = \sum_i V_n(|\vec{r}_i|) P_{n}(\hv{r}_i\cdot\hv{F})
\vspace{-3pt}
\end{equation}
which runs over all sites of the lattice except the origin. 
Consider the interaction energy for the dipole--dipole interaction $V=C_3 P_2(\hv{r}\cdot\hv{F})/r^3$ 
in a three-dimensional finite cubic lattice. 
Referring to Table \ref{table-1}, there is no cubic harmonic for $\ell=2$, and thus the
interaction energy $E$ vanishes by symmetry. 
The case of an infinite cubic lattice with dipole interactions
requires special treatment, since 
the result depends on how the sum is ordered (it is only conditionally convergent). 
This problem is discussed in Ref. \cite{Nijboer}, it has an interesting history going back to Lorentz.


As another illustration, consider an infinite two-dimensional square lattice in the $xy$ plane with lattice constant $a$.
The particles interact by quadrupole--quadrupole interactions $V=(C_5/P_4(0)) P_4(\hv{r}\cdot\hv{F})/r^5$. 
The interaction energy takes the form (cf.\ Table \ref{table-1} with symmetry $D_{4h}$):
\vspace{-6pt}
\begin{equation}
E = \frac{C_5}{a^5} \left[ P_4(\hv{F}\cdot\hv{z}) S_{5,0} + \sqrt{\frac{35\pi}{72}} \frac{Y^C_{44}(\hv{F})}{P_4(0)} S_{5,4}  \right]
\vspace{-3pt}
\end{equation}
with
\vspace{-6pt}
\begin{equation}
 S_{m,n} =  \sum_{p,q}{}' \frac{\cos(n \phi)}{(p^2+q^2)^{m/2}} ,
\vspace{-3pt}
\end{equation}
where $\phi$ is the polar angle of the point $(p,q)$ in the plane, and
the sum runs over all integers except the point $(0,0)$. The values of the sums are
$S_{5,0} = 5.09$ and $S_{5,4}=3.37$. 

\vspace{-6pt}
\section{Summary and Conclusions}
\vspace{-6pt}

In summary, we have investigated cold gases with anisotropic interparticle interactions.
We have discussed how to write down the interaction in a general way, 
with few assumptions on the form for the interparticle interactions. 
A few mathematical formulas were given which are useful to study such interactions.
The calculation of expectation values based on symmetry for such interactions was discussed
and illustrated by experimentally relevant cases. 
The effective interaction in a quasi two-dimensional geometry was calculated 
for a general class of potentials, it was found to be simpler than the special cases that exist
in the literature.

The point of view taken here is that it is profitable to study interactions with a general angular dependence.
It was found that such a general formulation can give simpler results than special cases (e.g.\ dipole--dipole),
as shown by discussion of symmetries and the calculation of the effective two-dimensional potential.

Future work will be to investigate the two-particle properties of anisotropic interactions,
i.e. bound states and scattering properties as pioneered in Ref. \cite{OMalley1964}.
Also, it would be interesting to study how anisotropic forces affect the many-body phases
of a system and result in novel quantum many-body phases.

\vspace{-6pt}
\begin{acknowledgments}
\vspace{-6pt}

Helpful discussions with 
C.~Becker, A.~Itin, A.~M.~Rey, K.~Sengstock, and M.~L.~Wall
are acknowledged.
The author was in part supported by the Hamburg Centre for Ultrafast Imaging.
\end{acknowledgments}

\appendix
\section{Evaluation of an integral involving spherical harmonics}

The purpose of this appendix is to evaluate the integral ($0 \le n $, $0 \le m \le n$ integers)
\begin{equation} 
\begin{split}
I_n^m = \frac{u}{\sqrt{2\pi}} \int_{-\infty}^\infty \! 
\frac{  \exp(-\frac{t^2 u^2}{2}) }{(1+t^2)^{(n+1)/2}} 
P_n^m\Bigl(\frac{t}{\sbsqrt{1+t^2}}\Bigr) \, dt .
\end{split}
\end{equation}
Here $P_n^m(x)$ are the associated Legendre polynomials (see e.g.\ \cite{Prudnikov-3}), which 
are related to the spherical harmonics by
$Y_{\ell m}(\theta,\phi)=\sqrt{\frac{(2\ell+1)(\ell-m)!}{4\pi(l+m)!}} P_\ell^m(\cos\theta) e^{i m \phi}$.
Since $P_n^m(-x)=(-1)^{n+m}P_n^m(x)$, the integral vanishes when $n+m$ is odd, and
in the following we restrict attention to the case where $n+m$ is even or zero.
We have then
\vspace{-6pt}
\begin{equation}\label{I-1}
 I_n^m = \frac{u}{\sqrt{2\pi}} \int_0^\infty \!\!
\frac{  \exp(-\frac{x u^2}{2}) }{(1+x)^{(n+1)/2}} 
P_n^m\Bigl(\!\sqrt{\frac{x}{1+x}}\Bigr) \, \frac{dx}{\sqrt{x}} .
\vspace{-3pt}
\end{equation}
It is possible to show that, for $n+m$ an even integer or zero, the following representation 
in terms of a Meijer G-function holds:
\vspace{-6pt}
\begin{equation}\label{I-2}
\frac{P_n^m \Bigl(\! \sqrt{\frac{x}{1+x}} \Bigr) }{(1+x)^{\frac{n+1}{2}}} =
\frac{2^n (-1)^{\frac{n+m}{2}}}{(n-m)!\,\sqrt{\pi}}
\MeijerG{12}{22}{x}{\frac{1-n+m}{2}, \frac{1-n-m}{2}}{\,0, \;\frac{1}{2}}
\end{equation}
% ...
Also, one has the integral
\vspace{-6pt}
\begin{align}\label{I-3}
 & \int_0^\infty \!\! x^{-1/2} e^{-c x} \MeijerG{12}{22}{x}{\frac{1-n+m}{2}, \frac{1-n-m}{2}}{\,0, \;\frac{1}{2}}
dx \nonumber\\[2pt]
&\: = \MeijerG{21}{12}{c}{\,\frac{1}{2}}{\frac{n-m}{2}, \frac{n+m}{2}} \nonumber\\[2pt]
&\: = c^\frac{n+m}{2} \Gamma({\tfrac{1-m+n}{2}})\Gamma({\tfrac{1+m+n}{2}}) \Psi\bigl(\tfrac{1+m+n}{2},1+m,c)
\vspace{-3pt}
\end{align}
where $\Psi$ is the Tricomi confluent hypergeometric function (often denoted by $U$).
In Eq. \refe{I-3} we have used a known integral for the Meijer G-function and simplified the result using 
some identities, with the help of Ref.\ \cite{Prudnikov-3}.
Now, inserting Eq. \refe{I-2} in Eq. \refe{I-1}, using Eq. \refe{I-3}, and simplifying the result one obtains:
\vspace{-6pt}
\begin{equation}
\! I_n^m = P_n^m(0) \, \Bigl(\frac{u^2}{2}\Bigr)^\frac{n+m+1}{2} \Psi\Bigl(\frac{n+m+1}{2},1+m,\frac{u^2}{2}\Bigr)
\vspace{-3pt}
\end{equation}
for $n+m$ an even integer or zero.

\end{document}